\documentclass[
    aps,      % American Physical Society journal
    prx,      % PRX Quantum
    reprint,  % approximate final formatting
    noeprint, % exclude eprints from journal references
    nofootinbib, % footnotes at bottom of page, instead of in bib
    superscriptaddress, % list all authors on same line
]{revtex4-2}
\usepackage{physics, siunitx, mathtools, amssymb, bm}
\usepackage[version=4]{mhchem}
\usepackage{graphicx}
\graphicspath{{./graphics/}}
\usepackage{hyperref}
\usepackage[capitalize]{cleveref}
\DeclareMathOperator*{\argmax}{arg\,max}
\DeclareMathOperator*{\argmin}{arg\,min}

\newcommand{\Vg}{V_\text{g}}
\newcommand{\Ic}{I_\text{c}}
\newcommand{\Icp}{I_{\text{c}+}}
\newcommand{\Icm}{I_{\text{c}-}}
\newcommand{\Icpm}{I_{\text{c}\pm}}

\begin{document}
\title{Gate-tunable polarity inversions and three-fold rotation symmetry of the superconducting diode effect}

\author{William F. Schiela}
\email{william.schiela@nyu.edu}
\author{Melissa Mikalsen}
\affiliation{Center for Quantum Information Physics, Department of Physics, New York University, New York, NY 10003, USA}

\author{Daniel Crawford}
\author{Stefan Ili\'c}
\affiliation{Department of Physics and Nanoscience Center, University of Jyv\"askyl\"a, P.O. Box 35 (YFL), FI-40014 University of Jyv\"askyl\"a, Finland}

\author{William M. Strickland}
\affiliation{Center for Quantum Information Physics, Department of Physics, New York University, New York, NY 10003, USA}

\author{F. Sebastian Bergeret}
\affiliation{Centro de F\'isica de Materiales (CFM-MPC), Centro Mixto CSIC-UPV/EHU, 20018 San Sebasti\'an, Spain}
\affiliation{Donostia International Physics Center (DIPC), 20018 Donostia-San Sebasti\'an, Spain}

\author{Javad Shabani}
\email{jshabani@nyu.edu}
\affiliation{Center for Quantum Information Physics, Department of Physics, New York University, New York, NY 10003, USA}

\date{\today}
\begin{abstract}
    The superconducting diode effect is an asymmetry in the critical current with respect to the supercurrent polarity.
    One impetus driving recent interest in the effect is its dependence on intrinsic or microscopic symmetry breaking mechanisms.
    Here, we study the superconducting diode effect in gated planar Josephson junctions fabricated on a superconductor--semiconductor heterostructure under an in-plane magnetic field.
    We observe two gate-driven inversions of the diode polarity in the vicinity of zero field, as well as a third-harmonic component in the dependence of the diode efficiency on the in-plane field angle.
    We analyze the Lifshitz invariant for an arbitrary spin--orbit coupling and show that multiple polarity inversions are possible in the presence of both linear and cubic Dresselhaus terms, where the Rashba parameter varies monotonically with gate voltage.
    Numerical calculations of the diode efficiency further reveal the presence of higher harmonics in its field-angle dependence in the presence of spin--orbit coupling.
\end{abstract}
\maketitle

\section{%
    \label{sec:introduction}%
    Introduction%
}

Non-reciprocal effects in superconductors, in particular the superconducting diode effect (SDE), are a field of intense research \cite{wakatsuki2017,yasuda2019nonreciprocal,ilic&bergeret2022,daido_intrinsic_2022,ando2020,he2022phenomenological,kokkeler2022,fominov2022asymmetric,ilic_superconducting_2024,alidoust2021,
jeon2022,kim2024intrinsic}.
The SDE, i.e.~the dependence of the critical current on the current direction, is a consequence of breaking both time-reversal and inversion symmetries.
Whereas time-reversal symmetry can be broken by an external magnetic field, inversion symmetry can be broken at different spatial scales \cite{kokkeler2024nonreciprocal}.  As a result, the SDE has been  observed in a wide range of materials and structures \cite{hou2023ubiquitous,souto&leijnse2022,cuozzo2024microwave}.
Of particular interest is the study of the SDE in systems with intrinsic inversion symmetry breaking, as it sheds light on the microscopic properties of the corresponding materials.

In this work, we study the SDE in planar Josephson junctions made of InAs with Al electrodes. We explore its symmetry with respect to the direction of the applied magnetic field, from which we uncover the role of different types of spin-orbit coupling, Rashba and Dresselhaus \cite{ganichev&golub2014review,baumgartner2022_diodeDresselhaus}, on transport properties.  Using a top gate, we are able to tune the ratio between these two contributions, which crucially affects the SDE.
We contrast our experimental results with a simple symmetry-based model and infer that the observed multiple polarity inversions of the diode effect can be explained by including not only linear-in-momentum SOC terms but also cubic Dresselhaus contributions.
A full numerical analysis of the system confirms the presence of higher harmonics in the field-angle dependence.

Our findings are significant not only for the development of spintronic devices, such as nonballistic spin-field effect transistors \cite{schliemann2003}, but also for the broader field of superconducting spintronics \cite{linder&robinson2015review}.

\section{%
    \label{sec:device-intro}%
    Planar Josephson junctions%
}
\begin{figure*}
    \centering
    \includegraphics{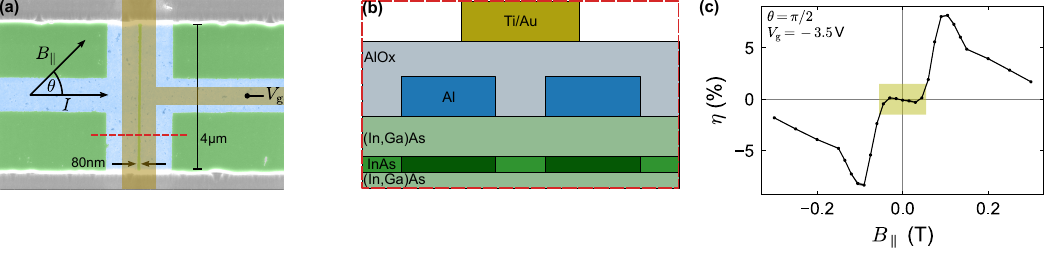}
    \caption{
        \textbf{(a)} Scanning electron micrograph of a representative device.  The superconductor (blue) and semiconductor (green) are falsely colored, and the gate (yellow) is shown schematically.  All devices are \SI{4}{\micro m} wide with \SI{80}{nm} separating the superconducting electrodes.  An in-plane magnetic field $B_\parallel$ is applied at an angle $\theta$ to the current bias $I$.  A gate voltage $\Vg$ is also applied.  The dashed red line shows the position of the cross section in (b).
        \textbf{(b)} Cross section of the heterostructure at the position marked in (a).  Not to scale.
        \textbf{(c)} Diode efficiency $\eta$ versus in-plane magnetic field $B_\parallel$ applied perpendicular to the current, $\theta=\pi/2$, at the gate voltage $\Vg=\SI{-3.5}{V}$.  The yellow shaded box highlights the feature of interest.
    }
    \label{fig:intro}
\end{figure*}

We study symmetric planar Josephson junctions in a superconductor--semiconductor heterostructure, as shown in \cref{fig:intro}(a).  The semiconductor consists of a near-surface InAs quantum well confining a two-dimensional electron gas with superconductivity proximity-induced by Al \citep{shabani2016,wickramasinghe2018}, as shown schematically in \cref{fig:intro}(b).  \Cref{fig:intro}(c) shows a representative measurement of the in-plane magnetic field dependence of the diode efficiency
\begin{equation}
    \eta=\frac{\Icp - \abs{\Icm}}{\Icp + \abs{\Icm}}
    \label{eq:diode-efficiency}
\end{equation}
in these devices.
When the in-plane field is perpendicular to the current, $\eta(B_\parallel)$ exhibits a roughly linear regime near zero field, followed by a sharp rise to a maximum.
(The diode efficiency is significantly suppressed when the field is parallel to the current; see
Fig.~S11.)
%\cref{si:fig:parperp}.)

Proposed mechanisms of the superconducting diode effect in planar Josephson junctions include Meissner screening in the superconducting contacts \citep{davydova2022}, orbital coupling to the in-plane magnetic field \citep{banerjee2023_diode}, and the interplay of spin--orbit coupling and Zeeman interaction \citep{yuan&fu2022}.
Our previous work \citep{schiela2025_diodeWsc} found a strong gate dependence at low field suggestive of a spin--orbit coupling mechanism at play.
%We expect orbital effects to be relatively weaker at low in-plane fields when the kinetic inductance of the thin Al is accounted for.
In the present work, we focus primarily on the diode efficiency at weak magnetic fields where an inversion of the diode polarity occurs at negative gate voltages, as highlighted in \cref{fig:intro}(c).

\section{%
    \label{sec:gate}%
    Gate-dependent diode polarity%
}
\begin{figure*}[tbp]
    \centering
    \includegraphics[]{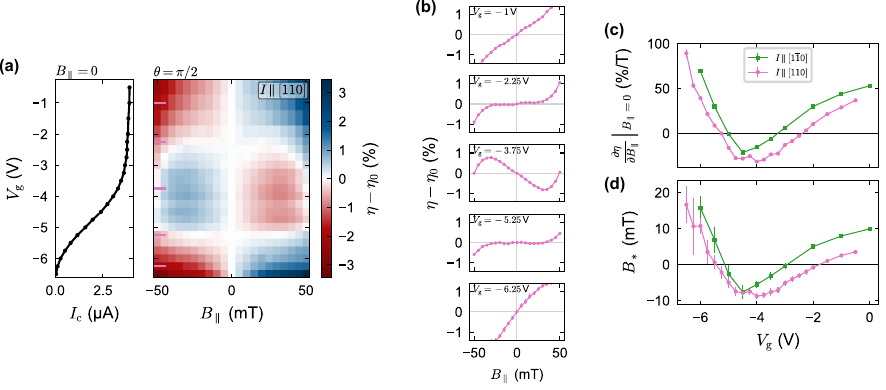}
    \caption{%
        \textbf{Gate-tunable diode polarity.}
        \textbf{(a)} Critical current $\Ic$ and diode efficiency $\eta$ versus gate voltage $\Vg$ and in-plane field $B_\parallel$ applied perpendicular to the current for a device with current along the $[110]$ crystal axis.  A gate-dependent offset $\eta_0 = \eta_0(\Vg, B_\parallel=0)$ has been subtracted; see
        Figs.~S1 and S7.
        %\cref{si:fig:zero-field-offset-tests,si:fig:slope:A1}.
        Pink ticks mark the line-cuts shown in (b).
        \textbf{(b)} Line-cuts of (a) at the gate voltages indicated.
        \textbf{(c)} Zero-field slope of the diode efficiency with respect to in-plane field for two mutually perpendicular devices.
        \textbf{(d)} In-plane field $B_*=\argmax_{B_\parallel}\Icp=-\argmin_{B_\parallel}\Icm$ (perpendicular to the current) at which the critical currents $\Icpm(B_\parallel)$ are extremized.
    }
    \label{fig:gate}
\end{figure*}

In \cref{fig:gate}(a) we show the gate voltage dependence of the critical current and diode efficiency at low in-plane magnetic fields $B_\parallel \lesssim \SI{50}{mT}$ applied perpendicular to the current.
We observe a change in the polarity of the diode efficiency in the regime where the critical current is suppressed by the gate voltage.
In contrast to Ref.~\citenum{shin2024}, this gate-tunable polarity inversion occurs in the vicinity of zero magnetic field.
\Cref{fig:gate}(b) shows line-cuts of the diode efficiency at constant gate voltage.
As the gate voltage is decreased, the zero-field slope of the diode efficiency with respect to field vanishes and then becomes negative.
Further decreasing the gate voltage causes the slope to vanish and become positive again.
Thus for any finite magnetic field less than \SI{50}{mT}, the diode polarity can be inverted by tuning the gate into the range \SI{-5.25}{V} to \SI{-2.25}{V}.
We note that the precise gate voltage values at which the diode polarity inverts are impacted by gate hysteresis; see
Fig.~S5.
%\cref{si:fig:hysteresis}.
Two polarity inversions are observed regardless of the crystallographic orientation of the device, as illustrated by the gate-dependent zero-field slope plotted in \cref{fig:gate}(c).
The zero-field slope is directly correlated with the field $B_*=\argmax_{B_\parallel}\Icp=-\argmin_{B_\parallel}\Icm$ at which the forward and backward critical currents $\Icpm$ are extremized, as shown in \cref{fig:gate}(d).

Within a model of the superconducting diode effect based on Rashba spin--orbit coupling \citep{lotfizadeh2024}, the sign of the extremal field $B_*$ is determined by the sign of the spin--orbit field.
The polarity inversions are then associated with changes in the helicity of the Rashba spin texture.
Changes in helicity can occur in semiconductor quantum wells if, for example, the electric field applied by the gate overcomes the built-in electric field in the valence band, or if the gate alters the charge distribution or band alignments in the heterostructure and thereby modifies the relative interface and intersubband contributions to the spin--orbit coupling \citep{fabian2007review,prager2021,farzaneh2024}.
%(see e.g.\ \citep[Fig.~III.1]{fabian2007review}).
The nonmonotonic behaviors of $B_*$ and $\eval{\pdv*{\eta}{B_\parallel}}_{B_\parallel=0}$ with gate voltage would then imply a nonmonotonic variation of the Rashba parameter $\alpha(\Vg)$.
% our recent weak antilocalization study has hinted at its possibility \citep{farzaneh2024}.
Measurements of weak antilocalization
(Figs.~S8 and S9)
%(\cref{si:fig:wal1,si:fig:wal2})
in a co-fabricated Hall bar indicate a nonmonotonic spin--orbit coupling strength but cannot determine its sign.
In \cref{sec:lifshitz}, we show that multiple polarity inversions with gate voltage are possible, even with \emph{monotonic} $\alpha(\Vg)$, if Dresselhaus spin--orbit couplings are additionally included.

\section{%
    \label{sec:angle}%
    Field angle dependence and three-fold symmetry%
}
\begin{figure*}[tbp]
    \centering
    \includegraphics[]{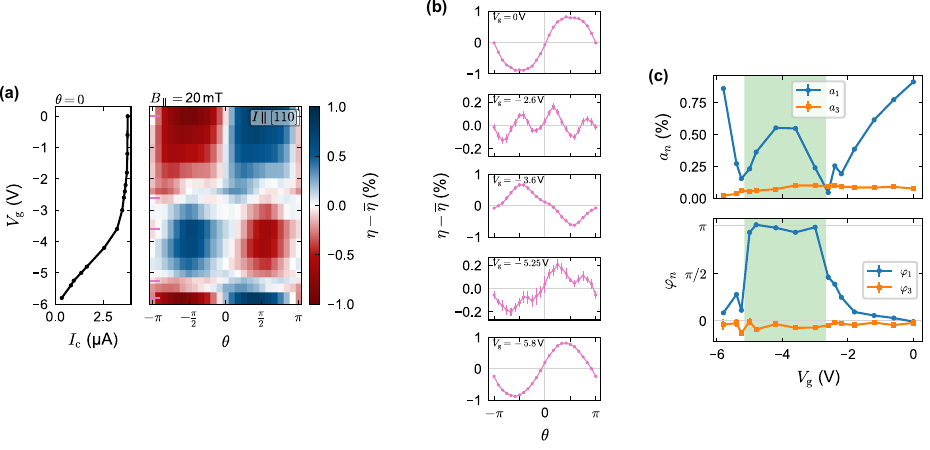}
    \caption{%
        \textbf{Field angle dependence and three-fold symmetry.}
        \textbf{(a)} Critical current $\Ic$ and diode efficiency $\eta$ versus gate voltage $\Vg$ and in-plane field angle $\theta$ with respect to current, at constant magnitude $B_\parallel=\SI{20}{mT}$, for a device with current $I$ along the $[110]$ crystal axis.  A gate-dependent mean $\overline{\eta} = \overline{\eta}(\Vg)$ has been subtracted (see 
        Fig.~S7).
        %\cref{si:fig:zero-field-offset-tests}).
        Pink ticks mark the line-cuts shown in (b).
        \textbf{(b)} Line-cuts of (a) at the gate voltages indicated.
        \textbf{(c)} Best fit parameters from fits of the data in (a) at each gate voltage to the model in \cref{eq:3theta-model}.  Fits are shown in
        Fig.~S10.
        %\cref{si:fig:anglefits}.
        The shaded region highlights the inverted polarity state.
    }%
    \label{fig:angle}
\end{figure*}

We now study the dependence of the diode efficiency on the in-plane magnetic field angle $\theta$ relative to the current, at constant $B_\parallel=\SI{20}{mT}$.
\Cref{fig:angle}(a) once again shows a gate-dependent inversion of the diode polarity coinciding with the sign changes of the zero-field slope shown in \cref{fig:gate}.
Line cuts in \cref{fig:angle}(b) reveal the presence of a third harmonic component of $\eta(\theta)$ that is particularly visible near the gate voltages where the polarity inversions occur and the fundamental component is suppressed.
We fit the angle dependence of the diode efficiency at each gate voltage to the model
\begin{equation}
    \eta(\theta) = \sum_{n \in \{1, 3\}} a_n \sin(n\theta + \varphi_n)
    \label{eq:3theta-model}
\end{equation}
including the fundamental $n=1$ and third harmonic $n=3$.
The best fit values of the amplitudes $a_n(\Vg)$ and phases $\varphi_n(\Vg)$ are shown in \cref{fig:angle}(c) as functions of gate voltage.
When the diode polarity inverts, the $n=1$ term vanishes and then changes sign, as evidenced by the vanishing and reemergence of the amplitude $a_1$ coinciding with a $\pi$-shift of the phase $\varphi_1$.
In contrast, the third harmonic term $n=3$ appears relatively insensitive to the gate voltage, maintaining a constant phase $\varphi_3\approx0$ and small amplitude $a_3$ that is suppressed at large negative voltages similarly to the critical current (\cref{fig:angle}(a)).
In \cref{sec:numerics}, we show numerically that higher harmonics in the angular dependence of the diode efficiency can arise in the free energy and are associated with higher-order terms in magnetic field.

\section{%
    \label{sec:theory}%
    Theory of the superconducting diode effect from spin--orbit coupling%
}
In this section, we present a theory that provides a qualitative explanation of some of the experimental observations regarding nonreciprocal transport.
\Cref{sec:lifshitz} presents a phenomenological analysis based on the Ginzburg-Landau (G-L) theory. Specifically, we write down the so-called Lifshitz invariant, the term in the free energy related to helical superconductivity \cite{bauer2012non,mineev1994helical,edelstein1996ginzburg,yip2002two,kaur2005helical,mineev2008nonuniform}, the anomalous phase\cite{buzdin2008,konschelle2015} and non-reciprocal effects, for an arbitrary spin-orbit coupling, and analyze it in different situations.
In \cref{sec:numerics}, we focus on a junction setup similar to the experimental configuration and solve the quasiclassical equations numerically, assuming a diffusive regime and arbitrary temperature.

Both methods are complementary: the phenomenological approach allows for the inclusion of any type of SOC, including cubic-in-momentum terms.  However, it focuses only on a ``bulk'' superconductor and on the linear response to the magnetic field and weak superconductivity.  In contrast, the numerical analysis focuses on the real Josephson junction and is valid for arbitrary temperatures and fields but considers only linear-in-momentum SOC.

\subsection{%
    \label{sec:lifshitz}%
    Lifshitz invariant for arbitrary spin--orbit couplings%
}
Magnetoelectric effects in superconductors, directly related to the appearance of an anomalous phase and non-reciprocal transport properties, are described by the so-called Lifshitz invariant in the free energy. In its most general form, this term can be written as:
\begin{equation}
F_L\propto {\mathbf T}.{\mathbf v}_s\; , \label{eq:Lif_inv}    
\end{equation}
where ${\mathbf v}_s$ is the superfluid velocity, proportional to the superconducting phase gradient ${\mathbf \nabla \phi}$, and ${\mathbf T}$ is a polar vector that is odd under time reversal. 

If we focus on effects linear in the Zeeman field $\mathbf{h}$ caused by an external magnetic field, the vector $\mathbf{T}$ can be written as \cite{bergeret2015theory}:
\begin{equation}
    T_j=h_i\gamma_{ij}\; ,
\end{equation}
where summation over repeated indices is implied. Here $\gamma_{ij}$ is a second-rank pseudo-tensor, allowed only in gyrotropic systems \cite{he2020magnetoelectric,kokkeler2024nonreciprocal}. 

Let us consider a generic Hamiltonian that describes a generic electronic system with spin-orbit coupling (SOC):
\begin{equation}
\mathcal{H}(\boldsymbol{p})=\frac{\boldsymbol{p}^2}{2m}-\mu + \boldsymbol{h}\cdot \boldsymbol{\sigma}+\boldsymbol{b}_{\boldsymbol{p}} \cdot \boldsymbol{\sigma}+V_{\rm imp}
\end{equation}
where $\boldsymbol{h}$ is the Zeeman field, and $\boldsymbol{b}_{\boldsymbol{p}}$ is SOC, with the property $\boldsymbol{b}_{-\boldsymbol{p}}=-\boldsymbol{b}_{\boldsymbol{p}}$. $V_{\rm imp}$ is a random spin-independent impurity potential, characterized by the scattering time $\tau$.

As shown in Ref. \cite{mal2005spin}, one can derive in the diffusive limit the following  second-rank pseudo tensor 
\begin{equation}
\gamma_{ij}=-4 \tau^2 \left\langle |\boldsymbol{b}_{\boldsymbol{p}}|^3 
\frac{\partial}{\partial p_j} \frac{b_{\boldsymbol{p}}^i}{|\boldsymbol{b}_{\boldsymbol{p}}|}
\right \rangle.\label{eq:gamma}
\end{equation}
where $\langle ...\rangle$ is the average over momentum direction. 
Substituting this into expression Eq. (\ref{eq:Lif_inv}) we arrive at the Lifshitz invariant for arbitrary SOC: 
\begin{equation}
F_L\sim  h_i\gamma_{ij}  \nabla_j \phi.\label{eq:LI_2}
\end{equation} 

Let us now consider a  specific system with Rashba spin-orbit coupling $(\alpha)$, and linear ($\beta_1$) and cubic ($\beta_3$) Dresselhaus coupling. The corresponding $\boldsymbol{b}$ vector is
\begin{align}
b_{\boldsymbol{p}}^x=(\beta_1+\alpha) p_y +2 \frac{\beta_3}{p_F^2} p_y (p_x^2-p_y^2) \nonumber \\
b_{\boldsymbol{p}}^y=(\beta_1-\alpha) p_x -2 \frac{\beta_3}{p_F^2} p_x (p_x^2-p_y^2).\label{eq:RD_SOC}
\end{align}
Here we have taken the $x$-direction along the [110]. Let us take that the current flows along the $x$ direction, so that $\nabla_x \phi\neq 0$ is finite, and that the Zeeman field $h$ is applied at an angle $\theta$ with respect to the current. From Eqs. (\ref{eq:gamma},\ref{eq:LI_2},\ref{eq:RD_SOC}) we obtain for the  Lifshitz invariant
\begin{multline}
F_L \sim h \sin \theta \big[(\alpha-\beta_1)(\alpha+\beta_1)^2 \\
-3 (\alpha-\beta_1)(\alpha+\beta_1)\beta_3+(6\alpha+2\beta_1)\beta_3^2-3\beta_3^3 \big].
\label{Eq:Lif}
\end{multline}
A similar expression also holds if the current is applied along the $[1\bar{1}0]$ direction with the substitution $\beta_1\to -\beta_1$ and $\beta_3 \to -\beta_3$. In the absence of cubic SOC, $F_L$ vanishes at $\alpha=\pm \beta_1$ --- this is the regime of persistent spin helix.

\begin{figure*}
    \centering
    \includegraphics[width= \textwidth]{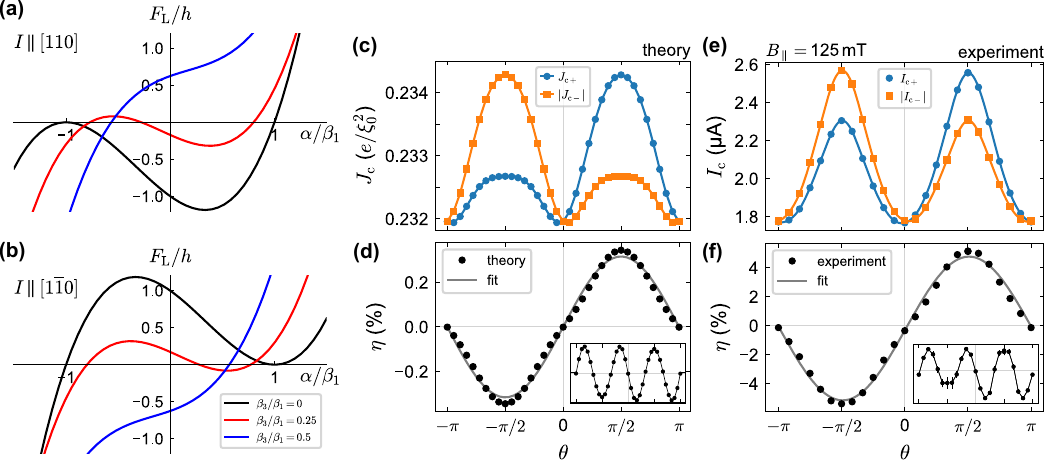}
    \caption{
        \textbf{(a,b)} Lifshitz invariant $F_\text{L}$ \eqref{Eq:Lif} as a function of the ratio $\alpha/\beta_1$ at different $\beta_3$ for current flowing along the \textbf{(a)} $[110]$ and \textbf{(b)} $[1\bar{1}0]$ directions.  The magnetic field $h$ is in both cases applied perpendicular to the current.
        \textbf{(c)} Positive and negative critical current densities $J_{\text{c}\pm}$ and \textbf{(d)} diode efficiency $\eta$ calculated numerically for $\alpha=2\Delta_0$, $\beta_1=1.1\Delta_0$, and $\beta_3=0$, as a function of in-plane magnetic field angle $\theta$.  The inset shows the residuals of the sinusoidal fit.
        \textbf{(e)} Measured positive and negative critical currents $\Icpm$ and \textbf{(f)} diode efficiency as a function of in-plane magnetic field angle at $B_\parallel=\SI{125}{mT}$, where the diode efficiency is maximized (see
        Fig.~S11),
        %\cref{si:fig:parperp}),
        and $\Vg=0$.  The inset shows the residuals of the sinusoidal fit.
        \label{fig:theory}
    }
\end{figure*}

By analyzing the Lifshitz invariant, we can estimate the diode efficiency as $\eta \sim F_L$. This is justified at low fields, so our phenomenological model can be compared to the zero-field slope data in our experiment and may explain the sign changes of $\eta$ that occur by tuning the gate voltage.  In \cref{fig:theory}(a,b) we plot \cref{Eq:Lif} for current applied along the $[110]$ and $[1\bar{1}0]$ crystal axes.  We see that in the absence of $\beta_3$ (black curve), $F_L$ and $\eta$ may change sign only once as Rashba SOC is tuned. Including moderate cubic SOC (red curve) enables more sign changes. However, if cubic SOC is too strong (blue curve), there is again only one sign change. In our experiment, the zero-field slope changes sign twice as the gate voltage is tuned for both directions of current. If we assume that $\alpha$ changes with the gate voltage in the experiment, while $\beta_1$ and $\beta_3$ remain mostly constant, we see that the case with finite moderate cubic SOC (red curve) captures best the experimental data.
Therefore, our phenomenological model suggests that the multiple sign changes of diode efficiency as a function of gate voltage may come from an interplay between linear Rashba, and linear and cubic Dresselhaus SOC.

\subsection{%
    \label{sec:numerics}%
    Numerical results%
}
In the phenomenological model presented above, we retain terms up to linear order in $h$ and obtain $\eta(\theta)\sim \sin \theta$.  Within this approximation, we cannot capture the third-harmonic contribution $\sin 3\theta$ to $\eta(\theta)$.  This would require keeping terms up to $h^3$, which would significantly complicate the simple analytical theory presented above.  However, third and higher harmonics in $\eta(\theta)$ are clearly seen in our microscopic simulations of an SNS junction with SOC presented in the following.

A generic superconducting system including SOC and magnetism is described by the Hamiltonian 
\begin{equation}
    H_{\rm SC} (\boldsymbol p, \boldsymbol r)= \tau_3 \mathcal H(\boldsymbol p, \boldsymbol r) + \hat \Delta(\boldsymbol r).
\end{equation}
Here $\hat\Delta(\boldsymbol r) = \tau_2 \sigma_2 \Delta (\boldsymbol r) e^{i \tau_3 \varphi}$ is the superconducting pair potential, and $\boldsymbol h, \boldsymbol{b_p}$ are now additionally functions of $\boldsymbol{r}$.
We solve the quasiclassical equations for this Hamiltonian in the diffusive regime using the methods and code described in Ref.~\citenum{virtanen_numeric_2024}.
We compute the diode efficiency $\eta$ for 2D SNS junctions with both Rashba and linear Dresselhaus SOC, and an in-plane exchange field.
The angle of the exchange field is defined relative to the direction of the Josephson current, as in the experiments:
\begin{equation}
\boldsymbol h (\boldsymbol r)  = (h(\boldsymbol r)  \cos \theta, h (\boldsymbol r) \sin \theta, 0).
\end{equation}
Currently, this method is limited to only linear-in-momentum SOC.
Thus we define $\boldsymbol {b_p}$ as in Eq.\,\ref{eq:RD_SOC}, but with $\beta_3 = 0$.
In all numerics we choose the following parameters: $(L_x, L_y, D, \sigma_{xy}', T)$ = $(2.5\xi_0, 2.5\xi_0, 1, 0.02, 0.1\Delta_0)$.
From \cite{ilic_superconducting_2024} we know that the length ($L_x$) of a junction is critical to realising the diode effect, while the width ($L_y$) is not so important.
Therefore, we choose a square junction to minimise the computational time of the numerics.

In \cref{fig:theory}(c,d) we compute the critical current and diode efficiency as a function of magnetic field angle
for a junction with $\alpha=2\Delta_0$ and $\beta_1=1.1\Delta_0$.
We observe that both the critical currents and diode effect are strongest when the magnetic field is perpendicular to the Josephson current, and the diode effect vanishes when the magnetic field is parallel to the Josephson current.
In addition, the diode efficiency contains a weak third harmonic, which can be seen in the residuals of the fit to a sinusoid containing only the first harmonic.
These features are consistent with the experimental data shown in \cref{fig:theory}(e,f).
In
Fig.~S12
%\cref{si:fig:numerics:field-angle}
we show numerical results for the cases where $\alpha=\beta_1$ and $\alpha<\beta_1$, both with $I\parallel[110]$.
For $\alpha=\beta_1$, SOC is effectively canceled and so there is no diode effect.
When $\alpha<\beta_1$, the diode efficiency remains strongest when the field is perpendicular to the current, but the polarity is inverted.
The critical currents, on the other hand, are maximum when the field is parallel to the current.
This is in contrast to the experimental data wherein the critical currents are always maximum when the field is perpendicular to the current, a phenomenon which has previously been attributed to flux focusing by the Al contacts \citep{suominen2017_fraunhofer} and is not described by the present model.

\section{%
    \label{sec:conclusion}%
    Conclusion%
}

We have studied critical current nonreciprocity, a facet of the superconducting diode effect, in symmetric planar Josephson junctions in a superconductor--semiconductor heterostructure comprising a near-surface InAs quantum well with superconducting correlations proximity-induced by Al.
We observed two inversions of the diode polarity near zero magnetic field by tuning gate voltage in two devices with mutually orthogonal orientations with respect to the underlying crystal system.
We further observed a third-harmonic component of the diode efficiency with respect to the angle of the in-plane magnetic field.
%These features may be explained by a nonmonotonic dependence of the Rashba spin-orbit coupling strength on gate voltage.
%Alternatively, 
We provided a phenomenological model of the diode effect incorporating linear Rashba, linear Dresselhaus, and cubic Dresselhaus spin--orbit couplings and showed that multiple polarity inversions can occur even when the gate-dependence of the Rashba parameter is monotonic.
Finally, we demonstrated the presence of higher harmonics in the diode efficiency by numerically calculating the free energy of the junction.
This work should be of interest to those studying nonreciprocal transport in systems with intrinsically or microscopically broken symmetries.

\section{Acknowledgments}
We thank Alex Matos-Abiague for fruitful discussions.  This work was supported by ONR N00014-22-1-2764 and ONR N00014-21-1-2450.  W.~F.~S. acknowledges support from the NDSEG Fellowship.

\bibliographystyle{apsrev4-2}
%\bibliography{bibs/papers_clean,bibs/textbooks_clean,bibs/theory}
\bibliography{bibs/merged}
\end{document}

\title{Supplementary Information: Gate-tunable polarity and three-fold rotation symmetry of the superconducting diode effect}

\author{William F. Schiela}
\email{william.schiela@nyu.edu}
\author{Melissa Mikalsen}
\affiliation{Center for Quantum Information Physics, Department of Physics, New York University, New York, NY 10003, USA}

\author{Daniel Crawford}
\author{Stefan Ili\'c}
\affiliation{Department of Physics and Nanoscience Center, University of Jyv\"askyl\"a, P.O. Box 35 (YFL), FI-40014 University of Jyv\"askyl\"a, Finland}

\author{William M. Strickland}
\affiliation{Center for Quantum Information Physics, Department of Physics, New York University, New York, NY 10003, USA}

\author{F. Sebastian Bergeret}
\affiliation{Centro de F\'isica de Materiales (CFM-MPC), Centro Mixto CSIC-UPV/EHU, 20018 San Sebasti\'an, Spain}
\affiliation{Donostia International Physics Center (DIPC), 20018 Donostia-San Sebasti\'an, Spain}

\author{Javad Shabani}
\email{jshabani@nyu.edu}
\affiliation{Center for Quantum Information Physics, Department of Physics, New York University, New York, NY 10003, USA}

\date{\today}
\maketitle

\begin{table}[htbp]
    \centering
    \begin{tabular}{cccc}
        \toprule
        Device & $\Wsc$ (\si{\micro{m}}) & $\hat{I}$ & data ID \\
        \midrule
        A & 0.9 & [110] & NNE \\
        B & 0.9 & $[1\overline{1}0]$ & ENE\\
        C & 0.6 & [110] & NNW \\
        D & 1.2 & $[1\overline{1}0]$ & E\\
        \bottomrule
    \end{tabular}
    \caption{Devices presented in this work along with their superconducting contact width $\Wsc$ and orientation with respect to the underlying crystal axes in terms of the direction of current $\hat{I}$.  The data ID is used in the included datasets.
    Data from Devices C and D are shown in \cref{fig:parperp} and \cref{main:fig:intro}(c), respectively.  All other data are from Devices A and B.}
    \label{tab:devices}
\end{table}

\begin{figure*}
    \centering
    \includegraphics{Wsc=0.9um-NNE_WFS06_slopes.pdf}
    \caption{
        \textbf{Device A, cooldown 6, $V_\text{g,init}=\SI{10}{V}$.} The sign of the current bias has been inverted to conform to the sign conventions defined in \cref{main:fig:intro}(a).
        The gate voltage was initialized to \SI{10}{V} (see \cref{fig:hysteresis}).
        \textbf{(a)} Diode efficiency $\eta=\left(\Icp-\abs{\Icm}\right)/\left(\Icp+\abs{\Icm}\right)$ as a function of gate voltage $\Vg$ and in-plane magnetic field $B_\parallel$ applied perpendicular to the current.
        \textbf{(b)} Gate dependence of the zero-field critical current $\Ic$.
        \textbf{(c)} Gate-dependent zero-field offset $\eval{\eta}_{B_\parallel=0}$ subtracted from the data in (a).
    }
    \label{fig:slope:A1}
\end{figure*}

\begin{figure*}
    \centering
    \includegraphics{Wsc=0.9um-NNE_WFS07_slopes.pdf}
    \caption{
        \textbf{Device A, cooldown 7, $V_\text{g,init}=\SI{10}{V}$.}
        The gate voltage was initialized to \SI{10}{V} (see \cref{fig:hysteresis}).
        \textbf{(a)} Diode efficiency $\eta=\left(\Icp-\abs{\Icm}\right)/\left(\Icp+\abs{\Icm}\right)$ as a function of gate voltage $\Vg$ and in-plane magnetic field $B_\parallel$ applied perpendicular to the current.
        \textbf{(b)} Gate dependence of the zero-field critical current $\Ic$.
        \textbf{(c)} Gate-dependent zero-field offset $\eval{\eta}_{B_\parallel=0}$ subtracted from the data in (a).
    }
    \label{fig:slope:A2}
\end{figure*}

\begin{figure*}
    \centering
    \includegraphics{Wsc=0.9um-ENE_WFS06_slopes.pdf}
    \caption{
        \textbf{Device B, cooldown 6, $V_\text{g,init}=\SI{0}{V}$.}
        The gate voltage was initialized to \SI{0}{V} (see \cref{fig:hysteresis}) before acquiring data from \SI{-3.1}{V} to \SI{-6.4}{V}, followed by \SI{-3.0}{V} to \SI{10}{V}.
        \textbf{(a)} Diode efficiency $\eta=\left(\Icp-\abs{\Icm}\right)/\left(\Icp+\abs{\Icm}\right)$ as a function of gate voltage $\Vg$ and in-plane magnetic field $B_\parallel$ applied perpendicular to the current.
        \textbf{(b)} Gate dependence of the zero-field critical current $\Ic$.
        \textbf{(c)} Gate-dependent zero-field offset $\eval{\eta}_{B_\parallel=0}$ subtracted from the data in (a).
    }
    \label{fig:slope:B1}
\end{figure*}

\begin{figure*}
    \centering
    \includegraphics{Wsc=0.9um-ENE_WFS06_slopes+10Vreset.pdf}
    \caption{
        \textbf{Device B, cooldown 6, $V_\text{g,init}=\SI{10}{V}$.}
        The gate voltage was initialized to \SI{10}{V} (see \cref{fig:hysteresis}) before recording data from \SI{0}{V} to \SI{-6}{V}.
        \textbf{(a)} Diode efficiency $\eta=\left(\Icp-\abs{\Icm}\right)/\left(\Icp+\abs{\Icm}\right)$ as a function of gate voltage $\Vg$ and in-plane magnetic field $B_\parallel$ applied perpendicular to the current.
        \textbf{(b)} Gate dependence of the zero-field critical current $\Ic$.
        \textbf{(c)} Gate-dependent zero-field offset $\eval{\eta}_{B_\parallel=0}$ subtracted from the data in (a).
    }
    \label{fig:slope:B2}
\end{figure*}

\begin{figure*}
    \centering
    \includegraphics{slopes-vs-vg+hysteresis.pdf}
    \caption{
        \textbf{Zero-field slope and gate hysteresis.}
        \textbf{(a)} Zero-field slope of the diode efficiency, $\eval{\dv{\eta}{B_\parallel}}_{B_\parallel=0}$, and \textbf{(b)} corresponding critical currents $\Ic$ as a function of gate voltage $\Vg$.  Two gate sweeps are shown for each of two devices, corresponding to the data in \cref{fig:slope:A1,fig:slope:A2,fig:slope:B1,fig:slope:B2}.  The gate voltage was initialized to $V_{\text{g},init}=\SI{10}{V}$ prior to all sweeps except one which was initialized to \SI{0}{V} (data at $\SI{-3.0}{V}\leq\Vg\leq\SI{10}{V}$ were acquired after the initial down-sweep for that particular series).
        The curves for Device A closely match, while those for Device B do not match due to the different initial gate voltages.
        \textbf{(c,d)} Repeated down- and up-sweeps of the gate voltage on $\Vg\in[\SI{-10}{V}, \SI{+10}{V}]$ for \textbf{(c)} Device A and \textbf{(d)} Device B.  Arrows indicate the sweep direction.  Initializing the gate voltage to $\pm\SI{10}{V}$ is sufficient to reliably select one of two hysteretic branches.
        The gate hysteresis can be understood in terms of the charge transfer dynamics between the quantum well, donor states in the (In,Al)As buffer, charge traps at the dielectric--semiconductor interface, and an inversion layer at that interface that occurs at high density \citep{prager2021}.
    }
    \label{fig:hysteresis}
\end{figure*}
% dependence of η on gate hysteresis, combined with prager2021 picture of changing charge distributions/band alignments/band bending, seem to also support SOC interpretation

\begin{figure*}
    \centering
    \includegraphics{bstar-vs-vg.pdf}
    \caption{In-plane magnetic field $B_*$ (perpendicular to the current) at which the critical current is maximized.  $B_*$ was extracted from fits of the $\Icpm(B_\parallel)$ data to the model of Ref.~\citenum{lotfizadeh2024}.  The first five $B_*(\Vg)$ traces shown here are from Ref.~\citenum{schiela2025_diodeWsc}.  The remaining four traces correspond to the data in \cref{fig:slope:A1,fig:slope:A2,fig:slope:B1,fig:slope:B2}.  Within the Rashba spin--orbit coupling model of Ref.~\citenum{lotfizadeh2024}, a change in the sign of $B_*$ implies a change in the sign of the spin--orbit field.}
    \label{fig:bstar}
\end{figure*}

\begin{figure*}
    \centering
    \includegraphics{zero-field-offset-tests.pdf}
    \caption{
        \textbf{Device A, zero-field offset tests.}
        \textbf{(a)} Comparison of various modifications of the measurement setup to a baseline measurement of the diode efficiency about zero magnetic field.  We compare the effects of swapping the polarities of the voltage probes and current source/sink and of removing the dc voltage amplifier in front of the voltmeter.
        Measurements were performed at $\Vg=\SI{-5.25}{V}$ where the zero-field slope of the diode efficiency in Device A is close to zero.
        \textbf{(b)} Measurements at $\Vg=\SI{-4.8}{V}$ where the zero-field slope of the diode efficiency in Device A is negative.
        The in-plane magnetic field was initially swept to $\SI{\pm500}{mT}$ to polarize any potential magnetic impurities that might be responsible for the zero-field offsets observed \citep{strambini2020}.
        In both (a) and (b), the magnitude of the zero-field offset varies slightly around \SI{-0.5}{\percent} and is always negative.
        Aside from this constant offset, the curves are antisymmetric, i.e.~$\eta(B_\parallel)-\eta_0=-\eta(-B_\parallel)+\eta_0$, where $\eta_0=\eval{\eta}_{B_\parallel=0}$.
        As shown in \cref{fig:slope:A1,fig:slope:A2,fig:slope:B1,fig:slope:B2}(b,c), the offset is always negative and peaked near the gate voltage where $\abs{\dv*{\Ic}{\Vg}}$ is maximum (n.b.~in \cref{fig:slope:A1} the current polarity has been inverted post-measurement).
        A finite diode effect at zero magnetic field requires some other source of broken time-reversal symmetry; however, we have not ruled out instrumental artifacts such as back-action of the supercurrent on the gate and cross-coupling between the gate voltage and current bias sources \citep{margineda2025}, which are particularly consistent with the correlation between $\abs{\eta_0}$ and $\abs{\dv*{\Ic}{\Vg}}$.
    }
    \label{fig:zero-field-offset-tests}
\end{figure*}

\begin{figure*}
    \centering
    \includegraphics{wal1.pdf}
    \caption{
        \textbf{Weak antilocalization in a co-fabricated Hall bar.}
        Average longitudinal conductance $\sigma_{(xx+yy)/2}=(\sigma_{xx}+\sigma_{yy})/2$ plotted relative to the mean, $\Delta\sigma=\sigma-\overline{\sigma}$ where $\overline{\sigma}$ is the field-averaged conductance for each trace, in units of the
        (normal) conductance quantum $e^2/h$, as a function of out-of-plane magnetic field $B_\perp$.
        Gate voltage $\Vg$ is stepped in \SI{0.1}{V} increments.
        The Hall bar was co-fabricated on the same chip as the junctions presented in this work.
    }
    \label{fig:wal1}
\end{figure*}

\begin{figure*}
    \centering
    \includegraphics{wal2.pdf}
    \caption{
        \textbf{Weak antilocalization in a co-fabricated Hall bar.}
        Symmetrized longitudinal conductance $\sigma_{jj}^\text{sym}=(\sigma_{jj}(B_\perp)+\sigma_{jj}(-B_\perp))/2$ for $j\in\{x, y\}$, plotted relative to the mean, $\Delta\sigma=\sigma-\overline{\sigma}$ where $\overline{\sigma}$ is the field-averaged conductance for each trace, in units of the (normal) conductance quantum $e^2/h$, as a function of out-of-plane magnetic field $B_\perp$.
        Gate voltage $\Vg$ is stepped in \SI{0.1}{V} increments.
        $x\parallel[110]$ and $y\parallel[1\overline{1}0]$.
        The Hall bar was co-fabricated on the same chip as the junctions presented in this work.
    }
    \label{fig:wal2}
\end{figure*}

\begin{figure*}
    \centering
    \includegraphics{anglefits.pdf}
    \caption{
        \textbf{Fits of the angular dependence.}
        Fits of the angle dependence data from Device A in \cref{main:fig:angle} to the model of \cref{main:eq:3theta-model}.  All angle sweeps started and ended at $\theta=\pi/2$, hence the two data points at that angle in each panel.  The data points at $\theta=\pi$ (open circles) are duplicated from the points at $\theta=-\pi$ and were not included in the fits.}
    \label{fig:anglefits}
\end{figure*}

\begin{figure*}
    \centering
    \includegraphics{parperp.pdf}
    \caption{
        \textbf{Device C, $V_\text{g}=\SI{0}{V}$.}
        Diode efficiency $\eta$ versus in-plane magnetic field $B_\parallel$ parallel ($\theta=0$) and perpendicular ($\theta=\pi/2$) to the current.
    }
    \label{fig:parperp}
\end{figure*}

\begin{figure*}
    \centering
    \includegraphics[width= \textwidth]{graphics/diode_effect_vs_angle.pdf}
    \caption{
        Critical currents $I_{c\pm}$ (a,c,e) and corresponding diode efficiency (b,d,e) for $\beta=1.1 \Delta_0$ and 
        (a,b) $\alpha=0$, 
        (c,d) $\alpha=1.1$, 
        (e,f) $\alpha=2\Delta_0$.
        For (b, f) dots indicate numeric result and solid line a sinusoidal fit. Insets indicate residuals.
        \label{fig:numerics:field-angle}
    }
\end{figure*}

\begin{figure*}
    \centering
    \includegraphics[width= 0.5 \textwidth]{graphics/h_vs_alpha.pdf}
    \caption{
        (a) Diode efficiency as a function of spin-orbit strength $\alpha$ and magnetic field strength $h$.
            The magnetic field is perpendicular to the Josephson current.
        (b) Zero-field gradient of the diode efficiency.
        Parameters: $\beta = 1.1\Delta$.
        \label{fig:numerics:field-magnitude}
    }
\end{figure*}

\bibliography{bibs/papers_clean}